# Mensen aanwijzen maar niet bij naam noemen: behavioural targeting, persoonsgegevens, en de nieuwe Privacyverordening[1]

## F.J. Zuiderveen Borgesius

In Europa is het gegevensbeschermingsrecht het belangrijkste juridische instrument om privacy te beschermen en een behoorlijke omgang met persoonsgegevens te bevorderen. Het gegevensbeschermingsrecht is alleen van toepassing als 'persoonsgegevens' verwerkt worden. Er is veel discussie over de vraag of het gegevensbeschermingsrecht van toepassing is als bedrijven gegevens over mensen verwerken maar daar geen naam aan koppelen. Zulke gegevens worden bijvoorbeeld gebruikt voor behavioural targeting. Bij deze marketingtechniek, een vorm van gepersonaliseerde communicatie, volgen bedrijven het gedrag van mensen op het internet, en gebruiken ze de verzamelde informatie om mensen gerichte advertenties te tonen. Deze bijdrage analyseert de discussie over de reikwijdte van het begrip 'persoonsgegeven', en trekt twee conclusies. Ten eerste blijkt uit een analyse van het geldende recht, in ieder geval volgens de interpretatie van de Europese privacytoezichthouders, dat de regels voor persoonsgegevens doorgaans van toepassing zijn op behavioural targeting. Ten tweede zouden die regels ook vanuit een normatief perspectief van toepassing moeten zijn.

---

## 1. Inleiding

In Europa is de Richtlijn Bescherming Persoonsgegevens[2] het belangrijkste instrument om privacy te beschermen en een behoorlijke omgang met persoonsgegevens te bevorderen. De richtlijn is alleen van toepassing als 'persoonsgegevens', gegevens over identificeerbare personen, verwerkt worden. Er is veel discussie over de vraag of de richtlijn van toepassing is als bedrijven gegevens over individuen verwerken maar daar geen naam aan koppelen. Zulke gegevens worden bijvoorbeeld gebruikt voor de marketingtechniek behavioural targeting. Bij behavioural targeting volgen bedrijven het online gedrag van mensen, en gebruiken ze de verzamelde informatie om mensen gerichte advertenties te tonen. Dit is een vorm van gepersonaliseerde communicatie.

Zijn de regels voor persoonsgegevens van toepassing op naamloze gegevens die verwerkt worden voor behavioural targeting, en zouden, vanuit het perspectief van fundamentele rechten, de regels van toepassing moeten zijn?

Paragraaf 2 geeft een inleiding over online marketing en behavioural targeting. Paragraaf 3 introduceert de regels voor persoonsgegevens. Deze bijdrage focust op de Europese regels; niet op de nationale implementatiewetten. Paragraaf 4 laat zien dat Europese privacytoezichthouders behavioural-targeting-gegevens doorgaans als persoonsgegevens beschouwen. De privacytoezichthouders zeggen dat een bedrijf dergelijke gegevens kan gebruiken om een persoon te individualiseren, 'to single out somebody', ook als het bedrijf geen naam kan koppelen aan de gegevens. Paragraaf 5 legt uit dat het vaak mogelijk is om een naam te koppelen aan naamloze gegevens. Paragraaf 6 bespreekt de nieuwe Verordening Persoonsgegevens, en de aparte regels daarin voor pseudonieme gegevens (dat zijn, kort gezegd, naamloze persoonsgegevens).

Dan neemt deze bijdrage een meer normatieve wending. Paragraaf 7 betoogt dat het gegevensbeschermingsrecht van toepassing zou moeten zijn op behavioural targeting. Paragraaf 8 bespreekt de tegenargumenten. Paragraaf 9 bevat twee conclusies. Ten eerste laat een analyse van het gegevensbeschermingsrecht zien, in ieder geval in de interpretatie van Europese privacytoezichthouders, dat het gegevensbeschermingsrecht doorgaans van toepassing is op behavioural targeting. Ten tweede zou dit recht ook vanuit een normatief perspectief van toepassing moeten zijn.

---

[2] Richtlijn 95/46/EG van het Europees Parlement en de Raad van de Europese Unie van 23 november 1995 betreffende de bescherming van natuurlijke personen in verband met de verwerking van persoonsgegevens en betreffende het vrije verkeer van die gegevens (*PbEG* 1995, L 281/31).

**2. Behavioural targeting**

Een groot deel van de gegevensverzameling via het internet is gerelateerd aan marketing. Bedrijven zoals Google en Facebook verzamelen informatie over honderden miljoenen mensen om gepersonaliseerde advertenties te tonen. Google bereikt naar eigen zeggen 90% van alle internetgebruikers,[3] en Facebook meldt dat het meer dan 1,5 miljard gebruikers heeft.[4] Maar minder bekende bedrijven beschikken ook over informatie over veel mensen, zoals The Rubicon Project (600 miljoen)[5] en Addthis (1,9 miljard)[6].

Veel bedrijven hopen dat onlineadvertenties effectiever zijn als ze precies aan de juiste persoon getoond worden. In een veelvoorkomend model voor reclame op het web betalen adverteerders een websitehouder alleen als op een advertentie geklikt wordt. Als duizend mensen een advertentie zien, klikken er gemiddeld minder dan twee mensen op die advertentie. In de branche wordt gesproken van een 'click-through rate' van 0,05 tot 0,2%.[7] Om dat percentage te verhogen is behavioural targeting ontwikkeld.

In een vereenvoudigd voorbeeld van behavioural targeting spelen drie partijen een rol: een internetgebruiker, een websitehouder, en een advertentienetwerk. Advertentienetwerken zijn bedrijven die advertenties tonen op duizenden websites. Zulke advertentienetwerken kunnen internetgebruikers volgen over die websites. Als iemand veel websites bezoekt over rechten, kan het advertentienetwerk afleiden dat die persoon geïnteresseerd is in dat vakgebied. Als die persoon een website bezoekt, kan het advertentienetwerk reclame tonen voor juridische boeken. Als iemand anders, die veel sites over koken bezoekt, gelijktijdig dezelfde website bezoekt, ziet hij mogelijk advertenties voor kookboeken.

---

[3] Google AdWords, 'About the Google Display Network', https://adwords.google.com/support/aw/bin/answer.py?hl=en&answer=57174. Alle websites in de noten zijn bezocht op 5 april 2016.

[4] Facebook schrijft: '1.59 billion monthly active users as of December 31, 2015', http://newsroom.fb.com/company-info/.

[5] http://rubiconproject.com/ad-buyers-solutions/.

[6] 'AddThis offers unparalleled insight into the interests and behaviors of over 1.9 billion web visitors', www.addthis.com/about.

[7] Zie bijvoorbeeld: D. Chaffey, 'Display advertising clickthrough rates', 21 april 2015: 'Across all ad formats and placements Ad CTR [clickthrough rate] is 0.06%', www.smartinsights.com/internet-advertising/internet-advertising-analytics/display-advertising-clickthrough-rates/.

Voor behavioural targeting wordt vaak gebruikgemaakt van http-cookies, kleine tekstbestandjes die een websitehouder opslaat (in de browser) op de computer van een bezoeker, om die computer later te herkennen. Deze cookies worden door veel websites gebruikt (*first party* cookies), bijvoorbeeld om de inhoud van een virtueel winkelwagentje te onthouden. Advertentienetwerken kunnen via de websites in hun netwerk ook cookies plaatsen op computers van websitebezoekers (*third party* cookies, of *tracking* cookies). Ieder tracking cookie heeft een unieke code, zoals '22be6e056ca010062||t=1392841778|cs=002213fd48e6bd6f7bf8d99065'.[8] Als iemand een andere website bezoekt waar het advertentienetwerk ook mee samenwerkt, kan het advertentienetwerk die persoon herkennen aan het cookie. Op deze manier kan het advertentienetwerk iemand volgen over alle websites waarop het advertenties toont. Via vrijwel elke populaire website worden tracking cookies geplaatst.

Naast dit soort http-cookies gebruiken bedrijven ook andere volgtechnieken, zoals flash cookies en andere super cookies, die meestal moeilijker te verwijderen zijn.[9] Ook wordt wel gebruikgemaakt van device fingerprinting: de meeste apparaten, zoals computers en smartphones, hebben unieke instellingen (browserversie, geïnstalleerde lettertypen, enz.) en kunnen op grond daarvan worden herkend.[10] Hoewel veel verschillende volgtechnieken worden gebruikt voor behavioural targeting, wordt in deze bijdrage verder kortheidshalve alleen over 'cookies' gesproken.

De gegevens die voor behavioural targeting worden verzameld kunnen betrekking hebben op vele online activiteiten: wat mensen lezen, welke video's zij kijken, wat zij zoeken, enz. Met deze informatie stellen bedrijven individuele profielen op over mensen. Zulke profielen kunnen verrijkt worden met locatiegegevens van gebruikers van *smartphones*, en andere gegevens die online en offline worden verzameld.

---

[8] Dit is de code van een cookie dat Google's dochterbedrijf DoubleClick op de computer van de auteur heeft geplaatst.

[9] Zie C.J. Hoofnagle e.a., 'Behavioral Advertising: The Offer You Cannot Refuse', (2012) 6(2) *Harvard Law & Policy Review* 273.

[10] G. Acar e.a., 'FPDetective: Dusting the web for fingerprinters', (2013) CCS '13 Proceedings of the 2013 ACM SIGSAC conference on Computer & Communications Security 1129.

Computerwetenschappers noemen individuele naamloze profielen niet anoniem, maar 'pseudoniem': 'A pseudonym is an identifier of a subject other than one of the subject's real names.'[11] Veel behavioural-targeting-bedrijven zeggen echter dat ze alleen 'anonieme' gegevens verwerken, en dat daarom de regels voor persoonsgegevens niet van toepassing zijn. Het Interactive Advertising Bureau, een brancheorganisatie voor online marketing, zegt over behavioural targeting:

'The information collected and used for this type of advertising is not personal, in that it does not identify you – the user – in the real world. No personal information, such as your name, address or email address, is used. Data about your browsing activity is collected and analysed anonymously.'[12]

In tegenstelling tot de interpretatie van het Interactive Advertising Bureau, zeggen Europese privacytoezichthouders dat gegevens die gebruikt worden om iemand te onderscheiden binnen een groep als persoonsgegevens beschouwd moeten worden – ook als bedrijven geen naam aan die gegevens koppelen.

### 3. Het recht op bescherming van persoonsgegevens

Privacy is een grondrecht dat wordt beschermd in verschillende verdragen, zoals het Europees Verdrag tot bescherming van de rechten van de mens. Het Handvest van de grondrechten van

---

[11] A. Pfitzmann & M. Hansen, 'A terminology for talking about privacy by data minimization: Anonymity, Unlinkability, Undetectability, Unobservability, Pseudonymity, and Identity Management' (Version v 0.34 10 augustus 2010), http://dud.inf.tu-dresden.de/Anon_Terminology.shtml, par. 9.

[12] Interactive Advertising Bureau Europe, 'Your Online Choices. A Guide to Online Behavioural Advertising, About', www.youronlinechoices.com/uk/about-behavioural-advertising. De tekst op de Nederlandse versie van de site is wat voorzichtiger: 'In de meeste gevallen zijn de gegevens die gebruikt worden voor het tonen van advertenties gebaseerd op jouw interesses geen persoonsgegevens, zoals je naam, adres of e-mailadres, en dus kun je niet worden geïdentificeerd. Het systeem is volledig geautomatiseerd. De informatie over jouw surfgedrag staat opgeslagen in een zogenaamde cookies [sic] (klein tekstbestandje) op jouw computer. Op basis van deze cookies wordt bepaald welke advertenties jij te zien krijgt, maar het is dus niet bekend wie jij bent', www.youronlinechoices.com/nl/veelgestelde-vragen.

de Europese Unie bevat, naast het recht op privacy, een apart recht op de bescherming van persoonsgegevens:

'Deze gegevens moeten eerlijk worden verwerkt, voor bepaalde doeleinden en met toestemming van de betrokkene of op basis van een andere gerechtvaardigde grondslag waarin de wet voorziet.'[13]

De Richtlijn Bescherming Persoonsgegevens (1995) geeft meer gedetailleerde regels. De richtlijn heeft onder meer als doel het beschermen van de fundamentele rechten en vrijheden, en vooral het recht op privacy. In Nederland is de richtlijn geïmplementeerd in de Wet bescherming persoonsgegevens.

De e-Privacyrichtlijn,[14] in Nederland geïmplementeerd in de Telecommunicatiewet, geeft een aparte regel voor het gebruik van cookies en dergelijke volgtechnieken.[15] Kort gezegd mogen bedrijven bepaalde cookies alleen plaatsen na toestemming van de internetgebruiker. De regels over cookies zeggen echter niets over het gebruik van gegevens die worden verzameld met cookies. De regels voor cookies blijven verder buiten beschouwing in deze bijdrage.[16] Voor zover er met cookies persoonsgegevens worden verwerkt, geldt het gegevensbeschermingsrecht uit de Richtlijn Bescherming Persoonsgegevens. Het gegevensbeschermingsrecht kent rechten toe aan mensen wier persoonsgegevens verwerkt worden ('betrokkenen'), en legt verplichtingen op aan partijen die persoonsgegevens verwerken ('verantwoordelijken', in deze bijdrage ook aangeduid als bedrijven).[17] Zo moeten bedrijven de door hun verwerkte persoonsgegevens beveiligen,[18] mogen zij geen disproportionele hoeveelheden persoonsgegevens verwerken,[19] en mogen zij persoonsgegevens niet langer bewaren dan noodzakelijk.[20] Betrokkenen hebben onder meer het recht om inzage te verkrijgen in de hen betreffende persoonsgegevens.[21]

---

[13] Artikel 7 Handvest van de grondrechten van de Europese Unie bevat het recht op privacy; artikel 8 het recht op bescherming van persoonsgegevens.

[14] Richtlijn 2002/58/EG, laatst gewijzigd door Richtlijn 2009/136/EG.

[15] Artikel 11.7a Telecommunicatiewet.

[16] Zie daarover: R. Leenes & E. Kosta, 'Taming the cookie monster with Dutch law – A tale of regulatory failure', 31 *Computer Law & Security Review* (2015), 317; F.J. Zuiderveen Borgesius, 'De nieuwe cookieregels: alwetende bedrijven en onwetende internetgebruikers?', *P&I* 2011, afl. 1, p. 2-11.

[17] Betrokkene: artikel 2 onderdeel a Richtlijn Bescherming Persoonsgegevens; verantwoordelijke: artikel 2 onderdeel d Richtlijn Bescherming Persoonsgegevens. De richtlijn onderscheid 'verwerkers' (artikel 2 onderdeel e) van verantwoordelijken; dat onderscheid valt buiten het bestek van deze bijdrage.

[18] Artikel 17 Richtlijn Bescherming Persoonsgegevens.

Zoals gezegd is de Richtlijn Bescherming Persoonsgegevens alleen van toepassing als 'persoonsgegevens' worden verwerkt. Vrijwel alles wat met persoonsgegevens gedaan kan worden, zoals verzamelen, opslaan, of analyseren, valt onder de definitie van 'verwerken'.[22] Maar zijn gegevens die verwerkt worden voor behavioural targeting 'persoonsgegevens'? De definitie van persoonsgegevens in de richtlijn luidt als volgt:

'"Persoonsgegevens", iedere informatie betreffende een geïdentificeerde of identificeerbare natuurlijke persoon, hierna "betrokkene" te noemen; als identificeerbaar wordt beschouwd een persoon die direct of indirect kan worden geïdentificeerd, met name aan de hand van een identificatienummer of van een of meer specifieke elementen die kenmerkend zijn voor zijn of haar fysieke, fysiologische, psychische, economische, culturele of sociale identiteit.'[23]

Persoonsgegevens zijn dus niet alleen gegevens zoals een naam en adres. Alle gegevens die betrekking hebben op een identificeerbare persoon zijn persoonsgegevens. Iemand is ook identificeerbaar als hij indirect kan worden geïdentificeerd. Het Hof van Justitie van de Europese Unie heeft meerdere malen bevestigd dat ook informatie zonder een naam een persoonsgegeven kan vormen.[24]

Op grond van het Handvest van de grondrechten van de Europese Unie moet een

---

[19] Artikel 6 lid 1 onderdeel c Richtlijn Bescherming Persoonsgegevens.

[20] Artikel 6 lid 1 onderdeel e Richtlijn Bescherming Persoonsgegevens.

[21] Artikel 12 Richtlijn Bescherming Persoonsgegevens.

[22] Artikel 2 onderdeel b Richtlijn Bescherming Persoonsgegevens.

[23] Artikel 2 onderdeel a Richtlijn Bescherming Persoonsgegevens.

[24] Zie bijvoorbeeld HvJ EU 24 november 2011, C-70/10, ECLI:EU:C:2011:771, r.o. 51 (*Scarlet/Sabam*). 'Aangezien die IP-adressen de precieze identificatie van die gebruikers mogelijk maken, vormen zij beschermde persoonsgegevens'; HvJ EU 11 december 2014, C-212/13, ECLI:EU:C:2014:2428, r.o.22 (*Ryneš*): 'Een door een camera vastgelegde afbeelding van een persoon valt derhalve onder het begrip persoonsgegevens in de zin van de in het vorige punt bedoelde bepaling, aangezien de betrokken persoon hierdoor kan worden geïdentificeerd.'

onafhankelijke toezichthouder toezien op de naleving van het gegevensbeschermingsrecht.[25] De Nederlandse privacytoezichthouder is de Autoriteit Persoonsgegevens (tot voor kort het College bescherming persoonsgegevens genaamd). De Europese privacytoezichthouders werken samen in een advies- en overlegorgaan met de naam Artikel 29-Werkgroep. De werkgroep heeft meer dan tweehonderd opinies gepubliceerd over de uitleg van het gegevensbeschermingsrecht.[26] De interpretatie van de werkgroep wordt vaak gevolgd door rechters en nationale privacytoezichthouders.[27] De opinies van de werkgroep zijn dus gezaghebbend; maar ze zijn niet juridisch bindend.

**4. Persoonsgegevens: 'to single out'**

In 2007 heeft de Artikel 29-Werkgroep een gedetailleerde opinie over het concept persoonsgegevens gepubliceerd.[28] De werkgroep merkt op dat vier elementen kunnen worden onderscheiden in de definitie van persoonsgegevens: (i) iedere informatie, (ii) betreffende, (iii) een geïdentificeerde of identificeerbare, (iv) natuurlijke persoon. Hieronder wordt behavioural targeting geanalyseerd aan de hand van deze vier elementen.

Element (i) – 'iedere informatie'. Persoonsgegevens die verwerkt worden voor behavioural targeting, zoals digitale informatie over iemands surfgedrag, vallen binnen de reikwijdte van 'iedere informatie'.[29]

Element (ii) – 'betreffende'. Bij behavioural targeting herkennen bedrijven een persoon vaak via een object, zoals zijn computer of telefoon. Jurisprudentie van het Hof van Justitie van de Europese Unie bevestigt dat informatie die verwijst naar een object een persoon kan identificeren.[30]

---

[25] Artikel 8 onderdeel 3 Handvest van de grondrechten van de Europese Unie.

[26] Op de website van de Artikel 29-Werkgroep zijn de opinies van de werkgroep te vinden: http://ec.europa.eu/justice/data-protection/article-29/index_en.htm.

[27] Zie over de werkgroep: S. Gutwirth & Y. Poullet, 'The contribution of the Article 29 Working Party to the construction of a harmonised European data protection system: an illustration of "reflexive governance"?', in: V.P. Asinari & P. Palazzi (red.), *Défis du Droit à la Protection de la Vie Privée. Challenges of Privacy and Data Protection Law*, Brussel: Bruylant 2008.

[28] Artikel 29-Werkgroep, Advies 4/2007 over het begrip 'persoonsgegeven' (WP 136), 20 juni 2007.

[29] Ibid., p. 6-9.

[30] In *Lindqvist* noemt het HvJ EG een telefoonnummer als een voorbeeld van informatie die iemand kan identificeren. Een telefoonnummer verwijst naar een apparaat (een telefoon), en niet direct naar een persoon.

De werkgroep legt uit dat informatie een persoon kan betreffen vanwege de inhoud, het resultaat, of het doel.[31] Informatie betreft een persoon vanwege de *inhoud*, als die 'over' een persoon gaat.[32] De werkgroep noemt een medisch dossier als voorbeeld. De gegevens in zo'n dossier betreffen duidelijk een persoon, ongeacht het doel of het gevolg van het gebruik van de gegevens. Als een bedrijf een individueel profiel voor behavioural targeting gebruikt, *betreft* dat profiel een persoon vanwege de *inhoud*. Over een persoon met cookie *xyz* op zijn of haar computer, kan een bedrijf bijvoorbeeld een lijst van bezochte websites, en een lijst van afgeleide interesses opslaan.

Behavioural-targeting-gegevens kunnen ook een persoon betreffen vanwege het *resultaat*.[33] Als een bedrijf een gerichte advertentie toont aan een bepaald persoon, behandelt het bedrijf die persoon anders dan anderen. Als een bedrijf een gerichte advertentie toont aan iemand, betreft de informatie in het individuele profiel die persoon vanwege het resultaat.

In 2014 oordeelde het Hof van Justitie van de Europese Unie dat een 'juridische analyse' in een dossier over een asielzoeker geen persoonsgegeven is.[34] Toch betekent dit arrest waarschijnlijk niet dat gegevens die betrekking hebben op een persoon vanwege het resultaat in het algemeen niet als persoonsgegevens beschouwd moeten worden. De uitspraak betreft een specifieke situatie: inzagerechten in dossiers in asielprocedures. De interpretatie van de werkgroep over gegevens die een persoon betreffen vanwege het resultaat wordt niet expliciet verworpen in het arrest.[35]

---

HvJ EG 6 november 2003, C-101/01, ECLI:EU:C:2003:596, r.o. 27 (*Lindqvist*) : '(…) op de eerste vraag [moet] worden geantwoord, dat het vermelden van verschillende personen op een internetpagina met hun naam of anderszins, bijvoorbeeld met hun telefoonnummer of informatie over hun werksituatie en hun liefhebberijen, als een "geheel of gedeeltelijk geautomatiseerde verwerking van persoonsgegevens" in de zin van artikel 3, lid 1, van richtlijn 95/46 is aan te merken.'

[31] Artikel 29-Werkgroep 2007, WP 136, p. 9-10.

[32] Ibid., p. 9.

[33] Ibid., p. 10-11.

[34] HvJ EU 17 juli 2014, gevoegde zaken C-141/12 en C-372/12, ECLI:EU:C:2014:2081 (*S. en M. en S./Minister voor Immigratie, Integratie en Asiel*).

[35] Zie E. Brouwer & F.J. Zuiderveen Borgesius, 'Access to Personal Data and the Right to Good Governance during Asylum Procedures after the CJEU's YS. and M. and S. judgment (C-141/12 en C-372/12)', *European Journal of Migration and Law* (2015)17, p. 259-272. Zie ook M. Jansen, 'Arrest HvJ EU inzake begrip persoonsgegevens en karakter inzagerecht', *P&I* 2014, afl. 5, p. 200-206, m.n. p. 205-206.

Informatie kan ook betrekking hebben op een persoon vanwege het *doel*. Volgens de Werkgroep is dit het geval als gegevens 'waarschijnlijk zullen worden gebruikt met *het doel* een persoon te beoordelen, op een bepaalde wijze te behandelen of de status of het gedrag van die persoon te beïnvloeden.'[36] Als een bedrijf een tracking cookie op een computer plaatst om gericht te adverteren, is het doel het beïnvloeden van een persoon – niet een apparaat. Sommige gegevensverwerkingen voor behavioural targeting betreffen geen persoonsgegevens. Een bedrijf kan persoonsgegevens gebruiken om een model op te stellen, bijvoorbeeld: *0,2% van de mensen die websites over sport bezoeken, klikt op advertenties voor sportschoenen, terwijl 0,1% van willekeurige mensen klikt op dergelijke advertentie.* Zo'n model verwijst niet naar een persoon en bestaat niet uit persoonsgegevens.[37]
Maar zodra zo'n model wordt toegepast op een persoon, *betreft* de informatie die persoon vanwege het *doel*. Als iemand met een cookie *xyz* op zijn computer bijvoorbeeld een website bezoekt, kan een advertentienetwerk die persoon (via het cookie) herkennen als iemand die veel websites bezoekt over sport. Het bedrijf heeft een model dat zegt dat mensen die websites over sport bezoeken relatief vaak klikken op advertenties voor sportschoenen. Daarom toont het bedrijf een advertentie voor sportschoenen. Op dat moment past het bedrijf het model toe op een bepaalde persoon, om hem te beïnvloeden.[38] Kortom, behavioural targeting leidt vaak tot de verwerking van gegevens die een persoon betreffen, vanwege de inhoud, het doel, of het resultaat van de verwerking.

Element (iii) – 'geïdentificeerd of identificeerbaar'. Brengt het verwerken van naamloze individuele profielen voor behavioural targeting met zich mee dat gegevens over 'identificeerbare' personen worden verwerkt?

---

[36] Artikel 29-Werkgroep 2007, WP 136, p. 11.

[37] Zo'n model wordt ook wel een groepsprofiel genoemd. Zie W. Schreurs e.a., 'Cogitas, Ergo Sum. The Role of Data Protection Law and Non-discrimination Law in Group Profiling in the Private Sector', in: M. Hildebrandt & S. Gutwirth (red.), *Profiling the European citizen: Cross-Disciplinary Perspectives*, New York: Springer 2008; B. Custers, *The Power of Knowledge, Ethical, Legal, and Technological Aspects of Data Mining and Group Profiling in Epidemiology*, Nijmegen: Wolf Legal Publishers 2004, p. 151; E. Siegel, *Predictive Analytics: The Power to Predict Who Will Click, Buy, Lie, or Die*, Hoboken: John Wiley & Sons 2013, p. 26.

[38] Zie B.J. Koops, 'Some reflections on profiling, power shifts, and protection paradigms', in: M. Hildebrandt & S. Gutwirth (red.), *Profiling the European citizen: cross-disciplinary perspectives*, New York: Springer 2008, p. 331; zie ook B. Custers, *The Power of Knowledge, Ethical, Legal, and Technological Aspects of Data Mining and Group Profiling in Epidemiology*, Nijmegen: Wolf Legal Publishers 2004, p. 151.

De definitie van persoonsgegevens uit de richtlijn noemt een 'identificatienummer' als een voorbeeld van informatie die een persoon kan identificeren. Tijdens het opstellen van de richtlijn in de jaren 90 van de vorige eeuw doelde de EU-wetgever waarschijnlijk op nummers zoals sofinummers. Maar unieke codes in cookies zijn ook lijsten van cijfers en letters, en kunnen als identificatienummers gezien worden.[39]

Een cookie of andere unieke *identifier* stelt een bedrijf in staat om iemands online gedrag te volgen, en een interesseprofiel van hem op te stellen. In de woorden van het Interactive Advertising Bureau: 'Cookies are used in behavioural advertising to identify users who share a particular interest so that they can be served more relevant adverts.'[40]

Volgens de werkgroep kan iemand geïdentificeerd worden zonder zijn naam te weten. In de opinie over persoonsgegevens uit 2007 zegt de werkgroep dat een 'naam *niet in alle gevallen noodzakelijk* is om een persoon te identificeren. Dit is het geval wanneer andere identificatiemiddelen worden gebruikt om iemand *van anderen te onderscheiden*'.[41] Een persoon is identificeerbaar als hij kan worden onderscheiden binnen een groep. In het Engels spreekt de werkgroep van 'to single out'.[42] Mogelijke vertalingen zijn onder meer 'individualiseren',[43] onderscheiden,[44] en 'isoleren'.[45] We zouden ook kunnen zeggen: iemand

---

[39] Zie C. Cuijpers, A. Roosendaal & B.J. Koops, 'D11.5: The legal framework for location-based services in Europe' (Future of Identity in the Information Society, FIDIS, 12 juni 2007), www.fidis.net/fileadmin/fidis/deliverables/fidis-WP11-del11.5-legal_framework_for_LBS.pdf, p. 25; P. Traung, 'EU Law on Spyware, Web Bugs, Cookies, etc. Revisited: Article 5 of the Directive on Privacy and Electronic Communications', (2010) 31 *Business Law Review* 216.

[40] Interactive Advertising Bureau, 'A Guide to online behavioural advertising' (Internet marketing handbook series) (2009), www.iabuk.net/sites/default/files/publication-download/OnlineBehaviouralAdvertisingHandbook_5455.pdf, p. 4.

[41] Artikel 29-Werkgroep 2007, WP 136, p. 14 (cursivering in origineel).

[42] Ibid. (Engelse versie), p. 14.

[43] Artikel 29-Werkgroep, 'Advies 5/2014 over anonimiseringstechnieken' (WP 216), 10 april 2014, p. 13.

[44] Artikel 29-Werkgroep, 'Advies 16/2011 over de Best Practice Recommendation on Online Behavioural Advertising van EASA/IAB' (WP 188), 8 december 2011, p. 9. Zie ook Artikel 29-Werkgroep, 'Advies 2/2010 over online reclame op basis van surfgedrag ("behavioural advertising")' (WP 171), 22 juni 2010; Artikel 29-Werkgroep, 'Working Document 02/2013 providing guidance on obtaining consent for cookies' (WP 208), 2 oktober 2013.

[45] G.J. Zwenne, *De verwaterde privacywet* (oratie Leiden), 12 april 2013, http://zwenneblog.weblog.leidenuniv.nl/files/2013/09/G-J.-Zwenne-De-verwaterde-privacywet-oratie-Leiden-

aanwijzen.

Volgens de werkgroep zou het een 'contradictio in terminis' zijn om te betogen dat iemand niet identificeerbaar is 'als het doel van de verwerking nu juist die identificatie is'.[46] In latere adviezen zegt de werkgroep expliciet dat cookies en soortgelijke bestanden met unieke identifiers persoonsgegevens zijn: 'such unique identifiers enable data subjects to be "singled out" for the purpose of tracking user behaviour while browsing on different websites and thus qualify as personal data.'[47]

Er zijn echter situaties denkbaar waarbij een tracking cookie niet een persoon betreft. Een computer in een internetcafé kan bijvoorbeeld door veel mensen worden gebruikt.[48] Een advertentienetwerk dat een profiel opstelt op basis van een cookie op die computer, stelt een profiel op op basis van het surfgedrag van een groep mensen. Een dergelijk profiel moet waarschijnlijk niet als een persoonsgegeven beschouwd worden.[49] Niettemin identificeren tracking cookies voor behavioural targeting meestal één persoon.[50]

Element (iv) – 'natuurlijk persoon'. Het vierde element van de persoonsgegevensdefinitie zegt dat informatie een 'natuurlijke persoon' moet betreffen.[51] Dit is meestal het geval met behavioural targeting.

Overigens is de juridische situatie in Nederland anders dan in de rest van de EU. De

---

12-apri-2013-NED.pdf, p. 13.

[46] Artikel 29-Werkgroep 2007, WP 136, p. 17.

[47] Artikel 29-Werkgroep, 'Opinion 16/2011 on EASA/IAB Best Practice Recommendation on Online Behavioural Advertising' (WP 188), 8 december 2011, p. 8.

[48] Dit voorbeeld komt uit Artikel 29-Werkgroep 2007, WP 136, p. 17. Zie ook Zwenne 2013, p. 10.

[49] Zie Zwenne 2013, p. 10-11.

[50] Het College bescherming persoonsgegevens suggereert (in de context van smart-tv's) dat gegevens die verzameld worden over een apparaat dat door meerdere mensen in één huishouden wordt gebruikt, doorgaans ook als persoonsgegevens gezien moeten worden (Onderzoek naar de verwerking van persoonsgegevens met of door een Philips smart tv door TP Vision Netherlands B.V., z2012-00605, Openbare versie Rapport definitieve bevindingen, juli 2013, p. 61-62).

[51] Een natuurlijk persoon is, kort gezegd, niet een rechtspersoon en niet een overledene. Zie over privacyrechten van overledenen: de speciale editie van *SCRIPTed* 2013, 10(1) en D. Korteweg & F.J. Zuiderveen Borgesius, 'E-mail na de dood. Juridische bescherming van privacybelangen', *P&I* 2009, afl. 5. Zie over de vraag of rechtspersonen beschermd worden door de richtlijn: Bart van der Sloot, 'Do privacy and data protection rules apply to legal persons and should they? A proposal for a two-tiered system', 31 *Computer Law & Security Review* (2015) 26.

Nederlandse Telecommunicatiewet bevat een rechtsvermoeden: kort gezegd wordt van het gebruik van tracking cookies vermoed dat het de verwerking van persoonsgegevens met zich meebrengt.[52]

Samenvattend: een analyse van het geldende recht, in ieder geval in de visie van Europese privacytoezichthouders, laat zien dat de regels voor persoonsgegevens doorgaans van toepassing zijn op behavioural targeting.

**5. Persoonsgegevens: bij naam noemen**

Afgezien van het feit dat iemand aanwijzen – *singling out somebody* – gezien kan worden als identificeren, is het vaak makkelijk om een naam te koppelen aan naamloze individuele profielen.

Een behavioural-targeting-bedrijf dat zulke naamloze profielen verwerkt, of een andere partij, kan vaak een naam aan de profielen koppelen. Ik onderscheid vier situaties waarin een bedrijf gegevens over een persoon verwerkt:

(i) Een bedrijf verwerkt gegevens over een persoon, en weet de naam van die persoon.

(ii) Een bedrijf verwerkt gegevens over een persoon, en het is vrij eenvoudig voor het bedrijf om een naam aan de gegevens te koppelen.

(iii) Een bedrijf verwerkt gegevens over een persoon, en het is moeilijk voor het bedrijf om een naam aan de gegevens te koppelen, maar het is vrij eenvoudig voor een andere partij om een naam aan de gegevens te koppelen.

(iv) Een bedrijf verwerkt gegevens over een persoon, en het is voor iedereen moeilijk om een naam aan de gegevens te koppelen.

Voor situatie (i) en (ii) is het niet controversieel dat het gegevensbeschermingsrecht van toepassing is. In situatie (i) is de persoon geïdentificeerd; in situatie (ii) is de persoon duidelijk identificeerbaar. Situatie (iii) leidt nog tot discussie; situatie (iv) des te meer.

---

[52] Artikel 11.7a Telecommunicatiewet luidt: 'Een handeling als bedoeld in het eerste lid [het plaatsen of uitlezen van cookies en vergelijkbare handelingen], die tot doel heeft gegevens over het gebruik van verschillende diensten van de informatiemaatschappij door de gebruiker of de abonnee te verzamelen, combineren of analyseren zodat de betrokken gebruiker of abonnee anders behandeld kan worden, wordt vermoed een verwerking van persoonsgegevens te zijn, als bedoeld in artikel 1, onderdeel b, van de Wet bescherming persoonsgegevens.'

Hieronder wordt elke situatie apart besproken.

*Situatie (i): een bedrijf weet de naam van de persoon*
In situatie (i) verwerkt een bedrijf gegevens over een persoon, en kent de naam van die persoon. Ter illustratie: Facebook, dat zijn gebruikers ook volgt op andere websites dan Facebook,[53] weet doorgaans de naam van die gebruikers.[54] Zo'n bedrijf verwerkt duidelijk gegevens over een geïdentificeerde persoon.

*Situatie (ii): een bedrijf kan een naam aan de gegevens koppelen*
In situatie (ii) verwerkt een bedrijf gegevens over een persoon, en is het vrij eenvoudig voor het bedrijf om een naam aan de gegevens te koppelen. In overweging 26 van de Richtlijn Bescherming Persoonsgegevens staat:

'Om te bepalen of een persoon identificeerbaar is, moet worden gekeken naar alle middelen waarvan mag worden aangenomen dat zij redelijkerwijs door degene die voor de verwerking verantwoordelijk is dan wel door enig ander persoon in te zetten zijn om genoemde persoon te identificeren; (…) de beschermingsbeginselen (zijn) niet van toepassing (…) op gegevens die op zodanige wijze anoniem zijn gemaakt dat de persoon waarop ze betrekking hebben niet meer identificeerbaar is.'[55]

De vraag is dus: welke middelen kan een verantwoordelijke (een bedrijf) *redelijkerwijs inzetten* om iemand te identificeren? Het antwoord hangt onder andere af van de stand van wetenschap en technologie, en van de kosten die gemoeid zijn bij het identificeren. De werkgroep zegt dat 'een slechts hypothetische mogelijkheid om iemand te onderscheiden niet voldoende is om die persoon als "identificeerbaar" te beschouwen'.[56]
Een behavioural-targeting-bedrijf kan vaak een naam koppelen aan gegevens over een persoon, gelet op 'alle middelen waarvan mag worden aangenomen dat zij redelijkerwijs door degene die voor de verwerking verantwoordelijk is dan wel door enig ander persoon in te zetten zijn' (overweging 26 van de richtlijn). Sommige behavioural-targeting-bedrijven leveren bijvoorbeeld diensten aan consumenten. Als een bedrijf een cookieprofiel van een persoon bezit, en een e-mailservice aan diezelfde persoon levert, kan het bedrijf het e-mailadres van de persoon koppelen aan het cookie. Veel e-mailadressen zijn persoonsgegevens, omdat zij iemand identificeren.[57] Bovendien bevatten e-mailadressen en e-mailberichten vaak de naam van de gebruiker. Ook een aanbieder van een sociale netwerksite of een andere dienst waarbij inloggegevens vereist zijn, kan de inloggegevens van de gebruiker koppelen aan het cookie op de computer van de betrokkene.

---

[53] G. Acar e.a., 'Facebook Tracking Through Social Plug-ins' (Technical report prepared for the Belgian Privacy Commission) (V. 1.1, 24 juni 2015), https://securehomes.esat.kuleuven.be/~gacar/fb_tracking/fb_plugins.pdf.
[54] Facebook's Name Policy www.facebook.com/help/292517374180078.
[55] Overweging 26 Richtlijn Bescherming Persoonsgegevens.
[56] Artikel 29-Werkgroep 2007, WP 136, p. 15.

Een zoekmachineaanbieder die een naamloos gebruikersprofiel met zoekopdrachten heeft over een persoon, kan ook vaak een naam koppelen aan het profiel. Het bedrijf zou veel mensen op basis van hun zoekopdrachten kunnen identificeren. Zoals een Google-medewerker zei in een rechtszaak: 'there are ways in which a search query alone may reveal personally identifying information'.[58] Als de gebruiker zijn eigen naam opzoekt, zou het nog makkelijker zijn voor de zoekmachineaanbieder om een naam aan het gebruikersprofiel te koppelen.[59] Kortom, een bedrijf dat naamloze individuele profielen verwerkt kan vaak vrij makkelijk een naam koppelen aan de gegevens. In zo'n geval verwerkt het bedrijf persoonsgegevens.

*Situatie (iii): een andere partij kan een naam aan de gegevens koppelen*

In situatie (iii) verwerkt een bedrijf gegevens over een persoon, en het is moeilijk voor het bedrijf om een naam toe te voegen aan de gegevens, maar het zou vrij makkelijk zijn voor een andere partij om dat te doen. Hoewel algemeen wordt erkend dat bedrijven in situatie (i) en (ii) persoonsgegevens verwerken, leidt situatie (iii) wel eens tot discussie.

Een hypothetisch voorbeeld van situatie (iii) is een advertentienetwerk, dat een cookieprofiel van een persoon heeft, gekoppeld aan een IP-adres. Laten we voor dit voorbeeld aannemen dat het moeilijk is voor het advertentienetwerk om een naam te koppelen aan het cookieprofiel. Maar de internet access provider van de persoon kan een naam koppelen aan het IP-adres. Die access provider wijst het IP-adres toe aan een abonnee, en heeft een overeenkomst met die abonnee. Een naam aan het IP-adres koppelen zou ook makkelijk zijn voor een onlinewinkel, als iemand een product bestelt en zijn naam en adres invult op de website van de winkel.[60]

---

[57] Een 'info@' e-mailadres verwijst vaak niet naar een persoon.

[58] M. Cutts, 'Declaration in Gonzales v. Google', 234 F.R.D. 674 (N.D. Cal. 2006) (17 februari 2006), http://docs.justia.com/cases/federal/district-courts/california/candce/5:2006mc80006/175448/14/0.pdf, p. 9.

[59] Zie C. Soghoian, 'The Problem of Anonymous Vanity Searches', I/S: A *Journal of Law & Policy for the Information Society* (2007) 3, 299.

Is het relevant dat alleen een andere partij een naam kan koppelen aan de gegevens? Nee, suggereert overweging 26 van de richtlijn: 'om te bepalen of een persoon identificeerbaar is, moet worden gekeken naar alle middelen waarvan mag worden aangenomen dat zij redelijkerwijs door degene die voor de verwerking verantwoordelijk is *dan wel door enig ander persoon* in te zetten zijn om genoemde persoon te identificeren (…)'.[61] De benadering van overweging 26 wordt wel de absolute benadering genoemd. Bij de absolute benadering betreffen gegevens een identificeerbaar persoon, als de verantwoordelijke (het bedrijf) de gegevens in handen heeft, en een *andere partij* de betrokkene kan identificeren aan de hand van die gegevens. Een relatieve benadering zou betekenen dat alleen wordt gekeken naar de middelen die ter beschikking staan van de *verantwoordelijke*, en niet naar de middelen die ter beschikking staan van derden.[62]

Overwegingen in richtlijnen hebben geen onafhankelijke juridische waarde, maar ze kunnen wel een dubbelzinnige bepaling uitleggen.[63] Bij de interpretatie van de Richtlijn Bescherming Persoonsgegevens kijkt het Hof van Justitie van de Europese Unie vaak naar de overwegingen.[64] Overweging 26 kan dus gebruikt worden om de definitie van persoonsgegevens te interpreteren.

Hoewel overweging 26 een absolute benadering suggereert, zijn de *middelen* ter beschikking van de verantwoordelijke (het bedrijf) soms wellicht toch relevant om te bepalen welke middelen redelijkerwijs gebruikt kunnen worden voor identificatie. Als menselijke haren gevonden worden door zomaar iemand, zijn die haren waarschijnlijk geen persoonsgegevens voor de vinder. Maar als de politie die haren naar een onderzoeksinstituut stuurt om ze te matchen aan DNA uit een databank met DNA van criminelen, moeten de haren waarschijnlijk als persoonsgegevens gekwalificeerd worden.[65]

---

[60] Websitehouders kunnen het IP-adres van websitebezoekers zien.

[61] Cursivering toegevoegd door de auteur.

[62] Zie European Commission's Information Society and Media Directorate-General, Legal analysis of a Single Market for the Information Society, chapter 4: The future of online privacy and data protection, prepared by DLA Piper 2011, http://ec.europa.eu/information_society/newsroom/cf/itemdetail.cfm?item_id=7022, p. 18-21; Zwenne 2013, p. 5-6.

[63] T. Klimas & J. Vaiciukaite, 'The Law of Recitals in European Community Legislation', ILSA *Journal of International & Comparative Law* (15) 2008, p. 2-33.

[64] Zie bijvoorbeeld HvJ EU 13 mei 2014, C-131/12, ECLI:EU:C:2014:317, r.o. 48, 54, 58, 66-67 (*Google Spain*); HvJ EG 6 november 2003, C-101/01, ECLI:EU:C:2003:596, r.o. 95 (*Lindqvist*).

Soms suggereerden privacytoezichthouders dat gegevens identificeerbaar waren voor één partij, terwijl de gegevens niet identificeerbaar waren voor een andere partij. Soms leken privacytoezichthouders dus te kijken naar de middelen van de verantwoordelijke (het bedrijf), en niet naar de middelen van anderen. De Engelse privacytoezichthouder leek bijvoorbeeld een dergelijke relatieve benadering van identificeerbaarheid te verdedigen.[66] Maar in een opinie uit 2014 kiest de werkgroep duidelijk de absolute benadering.[67]

Kortom, terwijl overweging 26 een absolute benadering lijkt voor te schrijven, kan de relatieve benadering relevant zijn bij het bepalen welke *middelen* redelijkerwijs gebruikt kunnen worden voor identificatie.

Er zijn verschillende manieren voor advertentienetwerken om een naam te koppelen aan gegevens. Zo lekken veel websites, vaak onbedoeld, namen of e-mailadressen van hun bezoekers aan advertentienetwerken. Sommige bedrijven zijn gespecialiseerd in het koppelen van namen aan cookieprofielen. Het doel van sommige web surveys – 'Win een gratis iPhone!' – is het koppelen van e-mailadressen aan cookieprofielen.[68] Een medewerker van

---

[65] Het voorbeeld is gebaseerd op Zwenne 2013, p. 6.

[66] Information Commissioner's Office, 'Anonymisation: Managing Data Protection Risk Code of Practice' (november 2012), http://ico.org.uk/for_organisations/data_protection/topic_guides/~/media/documents/library/Data_Protection/Practical_application/anonymisation-codev2.pdf, p. 21.

[67] Artikel 29-Werkgroep 2014, WP 216, p. 9: 'In de tweede plaats zijn "de middelen waarvan mag worden aangenomen dat zij redelijkerwijs in te zetten zijn (…) om genoemde persoon te identificeren" die welke worden gebruikt "door degene die voor de verwerking verantwoordelijk is dan wel door enig ander persoon". Het is daarom van wezenlijk belang in te zien dat wanneer een voor de verwerking verantwoordelijke de originele (identificeerbare) gegevens niet verwijdert op gebeurtenisniveau, en een deel van die dataset doorgeeft (bijvoorbeeld na het verwijderen of maskeren/afschermen van identificeerbare gegevens), de resulterende dataset nog steeds valt onder de noemer van persoonsgegevens.'

[68] Zie A. Narayanan, 'There is no such thing as anonymous online tracking' (Center for Internet and Society,

informatieleverancier Experian suggereerde in 2011 dat online profielen meestal aan een naam gekoppeld kunnen worden:

'Wij werken (…) samen met grote advertentienetwerken als Yahoo en MSN (…). Die beschikken over een netwerk van miljoenen websites, waarop zij alle bezoekers nauwkeurig in kaart brengen. Maar die profielen bevatten vaak geen of onvolledige NAW-gegevens. Onze algoritmes vinden meestal wel een aanknopingspunt om de verschillende profielen te matchen, bijvoorbeeld login-gegevens, en straks waarschijnlijk ook steeds vaker IP-adressen of een Facebook ID-nummer.'[69]

Als één bedrijf een e-mailadres aan een cookieprofiel heeft gekoppeld, kan het die informatie weer aan andere cookieprofielen koppelen ('cookie-synching').[70] Computerwetenschapper Narayanan vat samen: 'there is no such thing as anonymous online tracking'.[71] Bovendien is het vaak mogelijk om mensen binnen een ogenschijnlijk geanonimiseerde dataset te identificeren. In 2000 toonde Sweeney aan dat 87% van de Amerikaanse bevolking geïdentificeerd kon worden aan de hand van drie attributen: geboortedatum, geslacht en postcode.[72] Re-identificatietechnieken worden steeds beter.[73] Bovendien kan re-identificatie makkelijker worden als er meer datasets beschikbaar worden, bijvoorbeeld van sociale

---

Stanford Law School, 28 juli 2011), https://cyberlaw.stanford.edu/blog/2011/07/there-no-such-thing-anonymous-online-tracking.

[69] Geciteerd in A. Groot, 'Persoonsgegevens, het nieuwe goud', *Management Team*, 14 februari 2011, www.mt.nl/91/28332/finance/persoonsgegevens-het-nieuwe-goud.html.

[70] Het Interactive Advertising Bureau geeft deze omschrijving van cookie synching: 'a method of enabling data appending by linking one company's user identifier to another company's user identifier, to create a richer user profile at the cookie level', Interactive Advertising Bureau United States, 'Data Usage & Control Primer: Best Practices & Definitions' (2013) www.iab.net/media/file/IABDataPrimerv5.pdf.

[71] Narayanan 2011.

[72] L. Sweeney, 'Simple Demographics Often Identify People Uniquely, Data Privacy', Working Paper 3 Pittsburgh 2000, http://dataprivacylab.org/projects/identifiability/paper1.pdf. Zie ook L. Sweeney & J.S. Yoo, 'De-anonymizing South Korean Resident Registration Numbers Shared in Prescription Data', *Technology Science*, 2015, 092901, 29 september 2015, http://techscience.org/a/2015092901.

[73] Zie: M. Koot, *Measuring and Predicting Anonymity* (diss. Amsterdam UvA), 2012, https://cyberwar.nl/d/PhD-thesis_Measuring-and-Predicting-Anonymity_2012.pdf.

netwerksites, die kunnen worden gekoppeld aan de 'geanonimiseerde' data.[74]

Soms kan de persoon achter pseudonieme gegevens eenvoudig gevonden worden – zonder geavanceerde analyse. In 2006 publiceerde het Amerikaanse bedrijf AOL een dataset met individuele naamloze zoekprofielen van gebruikers van de AOL-zoekmachine. Binnen enkele dagen hadden journalisten van *The New York Times* de persoon achter zoekprofiel nr. 4.417.749 gevonden. Uit de zoekopdrachten bleek dat de zoeker waarschijnlijk een oudere vrouw met een hond was, uit de stad Lilburn. Een interview bevestigde dat zij zoekmachinegebruiker nr. 4.417.749 was.[75]

De discussie over naamloze behavioural-targeting-profielen doet denken aan de discussie over IP-adressen. De Artikel 29-Werkgroep en veel rechters in Europa zeggen dat IP-adressen in het algemeen moeten worden beschouwd als persoonsgegevens.[76] Anderen wijzen erop dat IP-adressen niet onder alle omstandigheden als persoonsgegevens beschouwd moeten worden.[77] Ten eerste verdedigen sommigen een relatieve benadering. Google suggereert bijvoorbeeld dat een IP-adres niet gezien moet worden als een persoonsgegeven, als een bedrijf zelf geen naam aan het adres kan koppelen (maar een andere partij wel).[78] Ten tweede kunnen IP-adressen soms niet worden gebruikt om een persoon te identificeren. In Qatar werd al het nationale internetverkeer door een paar IP-adressen gestuurd.[79] En veel organisaties, zoals de Universiteit van Amsterdam (waar ik werk), gebruiken één IP-adres voor al hun werknemers. In dergelijke gevallen is alleen een IP-adres zonder aanvullende gegevens doorgaans onvoldoende om iemand identificeren.

---

[74] C. Dwork, 'Differential Privacy: A Cryptographic Approach to Private Data Analysis', in: Julia Lane e.a. (red.), *Privacy, Big Data, and the Public Good: Frameworks for Engagement*, Cambridge: Cambridge University Press 2014, p. 297.

[75] M. Barbarom & T. Zeller, 'A Face Is Exposed for AOL Searcher No. 4417749', *The New York Times* 9 augustus 2006, www.nytimes.com/2006/08/09/technology/09aol.html.

[76] Zie Impact Assessment for the proposal for a Data Protection Regulation (SEC(2012)72 final, 25 januari 2012), Annex 2, p. 14-16; Time.lex, 'Study of case law on the circumstances in which IP addresses are considered personal data', *SMART* 2010/12 D3, Final report (mei 2011), www.timelex.eu/frontend/files/userfiles/files/publications/2011/IP_addresses_report_-_Final.pdf.

[77] Zie voor de meest gedetailleerde behandeling: Zwenne 2013.

[78] Zie A. Whitten, 'Are IP addresses personal?', Google Public Policy Blog, 22 februari 2008, http://googlepublicpolicy.blogspot.com/2008/02/are-ip-addresses-personal.html.

[79] J. Zittrain, *The Future of the Internet and How to Stop It*, New Haven: Yale University Press 2008, p. 157.

In *Scarlet/Sabam* (2011) oordeelt het Hof van Justitie van de Europese Unie dat IP-adressen in de handen van een internet access provider persoonsgegevens zijn.[80] De advocaat-generaal verwees naar opinies van de werkgroep om te onderbouwen dat IP-adressen als persoonsgegevens beschouwd moeten worden.

Toch is de discussie over IP-adressen nog niet voorbij. Het Hof was niet erg duidelijk in *Scarlet/Sabam*, maar het kan niet uitgesloten worden dat het Hof een relatieve benadering voor ogen had.[81] Voor partijen die geen internet access providers zijn is het moeilijker om een IP-adres aan een naam te koppelen. Zulke partijen zouden kunnen betogen dat IP-adressen in hun handen geen persoonsgegevens zijn. Maar in *Google Spain* zei de advocaat-generaal dat IP-adressen in de handen van Google persoonsgegevens zijn: een absolute benadering dus.[82] Het Hof heeft die conclusie bevestigd noch ontkracht. Binnenkort is er waarschijnlijk meer duidelijkheid. Een Duitse rechter heeft het Hof van Justitie van de Europese Unie gevraagd of IP-adressen als persoonsgegevens beschouwd moeten worden als zij niet in de handen zijn van een internet access provider.[83]

De jurisprudentie over IP-adressen bevestigt dat naamloze gegevens die verwijzen naar een apparaat persoonsgegevens kunnen zijn. Maar er is een belangrijk verschil tussen IP-adressen en behavioural-targeting-profielen. Behavioural-targeting-profielen bevatten meestal veel meer informatie dan een IP-adres. Daarom is het vaak makkelijker om een naam te koppelen aan een behavioural-targeting-profiel dan aan een IP-adres.[84] Verder kunnen behavioural-targeting-bedrijven doorgaans een IP-adres aan een cookieprofiel koppelen. Een bedrijf heeft normaliter het IP-adres van een computer of smartphone nodig om een advertentie naar dat apparaat te kunnen sturen.

---

[80] HvJ EU 24 november 2011, C-70/10, ECLI:EU:C:2011:771, r.o. 51 (*Scarlet/Sabam*).

[81] In een eerdere publicatie ging ik ervan uit dat het Hof alleen sprak over de IP-adressen in de handen van de access provider Scarlet. Bij nader inzien denk ik dat het Hof het mogelijk over IP-adressen in het algemeen had, omdat het spreekt over 'gebruikers' en niet over 'abonnees'. (Zie S. Kulk & F.J. Zuiderveen Borgesius, 'Filtering for Copyright Enforcement in Europe after the Sabam Cases', (2012) 34(11) *European Intellectual Property Review* 791-795.)

[82] Conclusie van A-G N. Jääskinen bij C-131/12, ECLI:EU:C:2013:424 (*Google Spain*), punt 30 en 48: 'Een aanbieder van een internetzoekmachine kan automatisch persoonsgegevens verzamelen die betrekking hebben op de gebruikers ervan, dat wil zeggen, op de personen die zoektermen invoeren in de zoekmachine. Deze automatisch overgebrachte gegevens kunnen bestaan uit hun IP-adres, gebruiksvoorkeuren (taal, enzovoort) (…)' (voetnoot verwijderd door auteur).

[83] C-582/14 (*Breyer*): 'Dient artikel 2, onder a), van richtlijn 95/46/EG (…) aldus te worden uitgelegd dat een internetprotocoladres (IP-adres) dat een aanbieder van diensten in verband met de toegang tot zijn internetsite opslaat, voor deze aanbieder reeds dan een persoonsgegeven vormt, wanneer een derde (in casu: de aanbieder van de toegang) beschikt over de bijkomende kennis die nodig is om de betrokken persoon te identificeren?'

Samenvattend: als een bedrijf naamloze individuele profielen verwerkt, en een andere partij kan een naam koppelen aan die profielen, volgt uit de preambule van de Richtlijn Bescherming Persoonsgegevens dat het bedrijf persoonsgegevens verwerkt.

*Situatie (iv): moeilijk voor bedrijf en anderen om een naam aan de gegevens koppelen*

In situatie (iv) verwerkt een bedrijf gegevens over een persoon, en is het voor iedereen moeilijk om een naam aan de gegevens te koppelen. Situatie (iv) is waarschijnlijk zeldzaam in de context van behavioural targeting. Het is immers vaak eenvoudig voor een behavioural-targeting-bedrijf om een naam te koppelen aan de gegevens die het verwerkt. Situatie (iv) werd besproken in paragraaf 4: het is niet doorslaggevend of een bedrijf een naam kan koppelen aan de gegevens. Als het bedrijf de gegevens gebruikt om iemand te onderscheiden binnen een groep – to single out somebody – verwerkt het bedrijf persoonsgegevens.

**6. Pseudonieme gegevens en de nieuwe Verordening Persoonsgegevens**

In januari 2012 presenteerde de Europese Commissie een voorstel voor de Verordening Bescherming Persoonsgegevens, die de Richtlijn Bescherming Persoonsgegevens uit 1995 moet vervangen.[85] Het voorstel voor de verordening heeft het debat over de reikwijdte van het begrip 'persoonsgegeven' aangewakkerd. Er is veel gelobbyd om het voorstel minder belastend voor het bedrijfsleven te maken.[86] Veel bedrijven zeiden dat pseudonieme gegevens, zoals naamloze behavioural-targeting-profielen, buiten de werkingssfeer van het regime zouden moeten vallen, of dat er een lichter regime zou moeten gelden voor

---

[84] Zie J. van Hoboken, *Search engine freedom: on the implications of the right to freedom of expression for the legal governance of search engines*, Alphen aan den Rijn: Kluwer Law International 2012, p. 328.

[85] European Commission, Proposal for a Regulation of the European Parliament and of the Council on the Protection of Individuals with regard to the Processing of Personal Data and on the Free Movement of Such Data (General Data Protection Regulation) COM(2012)11 final, 2012/0011(COD), 25 januari 2012.

[86] Zie I. Traynor e.a., '30,000 lobbyists and counting: is Brussels under corporate sway?', *The Guardian* 8 mei 2014, www.theguardian.com/world/2014/may/08/lobbyists-european-parliament-brussels-corporate.

pseudonieme gegevens.[87]

Het begrip 'persoonsgegeven' is vastgelegd in primair EU-recht (het Handvest van de grondrechten van de EU en het Verdrag betreffende de werking van de EU[88]). Hijmans merkt op dat de EU-wetgever daarom geen uitzondering (voor pseudonieme gegevens) op de werkingssfeer van het gegevensbeschermingsrecht kan vaststellen in secundaire EU-wetgeving.[89]

In 2014 heeft het Europees Parlement een aangepaste versie van de verordening aangenomen.[90] Die versie introduceerde een nieuwe categorie persoonsgegevens: 'pseudonieme gegevens'.[91] De regels waren minder streng voor dergelijke gegevens.[92]

In juni 2015 heeft de Europese Raad zijn voorstel gepubliceerd om op grond daarvan de onderhandelingen te beginnen.[93] Het voorstel van de Raad bevatte ongeveer dezelfde definitie

---

[87] Zie bijvoorbeeld: N. Stringer, IAB UK, 'Could "Pseudonymous Data" be the Compromise Where the Privacy Battle is Settled?', 14 maart 2013, www.exchangewire.com/blog/2013/03/14/iab-uk-could-pseudonymous-data-be-the-compromise-where-the-privacy-battle-is-settled/; Amazon EU Sarl, Proposed amendments to MEP Gallo's opinion on data protection, https://wiki.laquadrature.net/images/7/71/AMAZON-amendments.pdf; Yahoo! Rationale for Amendments to Draft Data Protection Regulation as Relate to Pseudonymous Data, www.centerfordigitaldemocracy.org/sites/default/files/Yahoo_on_Pseudonymous_Data-1.pdf.

[88] Artikel 16 Verdrag betreffende de werking van de Europese Unie (geconsolideerde versie); artikel 8 Handvest van de grondrechten van de Europese Unie.

[89] H. Hijmans, *The European Union as a Constitutional Guardian of Internet Privacy and Data Protection: the Story of Article 16 TFEU* (diss. Amsterdam UvA), 2016, http://dare.uva.nl/record/1/511969, p. 234 (hoofdstuk 6, par. 2).

[90] LIBE Compromis, proposal for a Data Protection Regulation. Deze bijdrage verwijst naar de Inofficial Consolidated Version after LIBE Committee Vote, provided by the Rapporteur, General Data Protection Regulation, 22 oktober 2013, www.janalbrecht.eu/fileadmin/material/Dokumente/DPR-Regulation-inofficial-consolidated-LIBE.pdf.

[91] Ibid., artikel 4 lid 2 onderdeel a.

[92] Ibid., artikel 2 onderdeel a, artikel 6 onderdeel f, overweging 38 en 58a. Bedrijven mogen volgens de preambule, onder bepaalde voorwaarden, pseudonieme gegevens gebruiken voor behavioural targeting, zonder voorafgaande toestemming van de betrokkene.

[93] Proposal for a Regulation of the European Parliament and of the Council on the protection of individuals with regard to the processing of personal data and on the free movement of such data (General Data Protection Regulation), Preparation of a general approach, Brussel, 11 juni 2015, 9565/15, 2012/0011(COD) http://data.consilium.europa.eu/doc/document/ST-9565-2015-INIT/en/pdf.

van persoonsgegevens als het voorstel van de Europese Commissie uit 2012.[94] Het voorstel van de Raad bevatte ook een definitie van pseudonimiseren, die grofweg overeenkwam met de definitie door het Europees Parlement.[95]

In reactie op het voorstel van de Raad, zei de Artikel 29-Werkgroep opnieuw dat 'single-out'-gegevens als persoonsgegevens beschouwd moeten worden.[96] De werkgroep was tegen de introductie van een apart regime voor pseudonieme gegevens.[97] Reding, toen nog eurocommissaris, had al gewaarschuwd: '(…) pseudonymous data must not become a Trojan horse at the heart of the Regulation, allowing the non-application of its provisions.'[98]

In december 2015 hebben het Parlement en de Raad een akkoord bereikt over de verordening.[99] Op het moment van schrijven moeten het Europees Parlement en de Raad deze 2015-tekst nog formeel aannemen. De 2015-tekst van de verordening omschrijft persoonsgegevens als volgt:

'"Personal data" means any information relating to an identified or identifiable natural person "data subject"; an identifiable person is one who can be identified, directly or indirectly, in particular by reference to an identifier such as a name, an identification number, *location data, online identifier* or to one or more factors specific to the physical, physiological,

---

[94] Ibid., artikel 4 lid 1.

[95] Ibid., artikel 4 lid 3 onderdeel b. Zie over pseudonieme data ook overweging 23, 23c, 60a, 61, 67, en artikel 23 lid 1, artikel 28 lid 4 onderdeel b, artikel 30 lid 1, artikel 31 lid 1, artikel 32 lid 1, artikel 33 lid 1, artikel 38 lid 1 onderdeel a en artikel 83 lid 2.

[96] Artikel 29-Werkgroep. 'Core topics in the view of trilogue' (17 juni 2015), http://ec.europa.eu/justice/data-protection/article-29/documentation/other-document/files/2015/20150617_appendix_core_issues_plenary_en.pdf, p. 5.

[97] Ibid., p. 6.

[98] V. Reding, 'The EU Data Protection Regulation: Promoting Technological Innovation and Safeguarding Citizens' Rights' (Speech, Intervention at the Justice Council, 4 maart 2014), http://europa.eu/rapid/press-release_SPEECH-14-175_en.htm?locale=en.

[99] Regulation (EU) No XXXX/2016 of the European Parliament and of the Council on the Protection of Individuals with Regard to the Processing of Personal Data and on the Free Movement of Such Data (General Data Protection Regulation), Consolidated text (outcome of the trilogue of 15/12/2015), www.janalbrecht.eu/fileadmin/material/Dokumente/GDPR_consolidated_LIBE-vote-2015-12-17.pdf; hierna: Verordening Bescherming Persoonsgegevens (2015-tekst).

genetic, mental, economic, cultural or social identity of that person.'[100]

Een van de belangrijkste verschillen met de definitie in de huidige richtlijn is dat nu 'location data' en 'online identifier' worden genoemd als voorbeelden van identifiers.[101] De preambule van de verordening kiest een absolute benadering ten aanzien van identificeerbaarheid, en noemt 'single out' als een manier om mensen te identificeren: 'To determine whether a person is identifiable, account should be taken of all the means reasonably likely to be used, *such as singling out, either by the controller or by any other person* to identify the individual directly or indirectly.'[102]

De 2015-tekst van de verordening bevat ook een definitie van pseudonimiseren:

'"Pseudonymisation" means the processing of personal data in such a way that the data can no longer be attributed to a specific data subject without the use of additional information, as long as such additional information is kept separately and subject to technical and organisational measures to ensure non-attribution to an identified or identifiable person.'[103]

De preambule bevestigt dat pseudonieme gegevens nog steeds persoonsgegevens kunnen zijn: 'Data which has undergone pseudonymisation, which could be attributed to a natural person by the use of additional information, should be considered as information on an identifiable natural person.'[104]

De 2015-tekst van de verordening behandelt pseudonimiseren vooral als een beveiligingsmaatregel.[105] Verder suggereert de verordening dat, onder bepaalde omstandigheden, bedrijven niet hoeven te voldoen aan inzageverzoeken als het gaat om

---

[100] Artikel 4 lid 1 Verordening Bescherming Persoonsgegevens (2015-tekst) (hoofdletter aangepast; cursivering toegevoegd).

[101] De woorden 'genetic' en 'economic' (identity) zijn ook nieuw in het artikel.

[102] Overweging 23 Verordening Bescherming Persoonsgegevens (2015-tekst).

[103] Artikel 4 lid 3 onderdeel b Verordening Bescherming Persoonsgegevens (2015-tekst) (hoofdletter aangepast).

[104] Overweging 23 Verordening Bescherming Persoonsgegevens (2015-tekst).

[105] Artikel 23 lid 1; artikel 30 lid 1 onderdeel d; artikel 83 onderdeel 1; overweging 60a, overweging 61, 67 en 125 Verordening Bescherming Persoonsgegevens (2015-tekst).

pseudonieme gegevens.[106] Een bespreking van dat artikel valt buiten het bestek van deze bijdrage.[107]

In de preambule van de versie van het Europees Parlement (2014) werd nog gesuggereerd dat bedrijven geen toestemming van de betrokkene hoefden te vragen voor behavioural targeting, zolang zij alleen pseudonieme gegevens gebruikten.[108] Deze suggestie is vervallen in de 2015-tekst.

Kortom, de Verordening Persoonsgegevens zegt expliciet dat pseudonieme gegevens persoonsgegevens kunnen zijn. De preambule noemt 'to single out somebody' als een voorbeeld van identificeren. Toch zal de discussie over de reikwijdte van het begrip 'persoonsgegeven' waarschijnlijk nog niet over zijn.[109]

## 7. Argumenten voor de 'single-out'-interpretatie

Veel (maar niet alle) auteurs zeggen dat behavioural-targeting-gegevens als persoonsgegevens beschouwd moeten worden.[110] Maar *waarom* moeten behavioural-targeting-gegevens als persoonsgegevens worden aangemerkt?

Ten eerste brengt de verwerking van pseudonieme gegevens voor behavioural targeting risico's met zich mee die het gegevensbeschermingsrecht nu juist probeert in te perken. De risico's bestaan ook als een profiel niet aan een naam, maar aan een andere indentifier wordt gekoppeld. Zo kan een *chilling effect* optreden als gevolg van grootschalige gegevensverzameling.[111] Mensen passen hun gedrag aan als zij weten dat ze worden geobserveerd. Het voelt ongemakkelijk om een website over kanker of een politieke partij te bezoeken als je weet dat je internetgebruik geanalyseerd kan worden door allerlei partijen.[112] Verder hebben mensen amper controle over hun persoonsgegevens. Mensen weten niet welke informatie over hen wordt verzameld, hoe die gebruikt wordt, en met wie die wordt gedeeld. Het gevoel geen controle te hebben over de eigen gegevens is op zichzelf al een privacyprobleem. Verder brengt grootschalige gegevensopslag risico's met zich mee. Zo kan een datalek optreden, of kunnen gegevens worden gebruikt voor onverwachte doeleinden. Geheime diensten gebruiken soms door bedrijven geplaatste tracking cookies om mensen te identificeren.[113]

---

[106] Artikel 10 Verordening Bescherming Persoonsgegevens. Zie ook overweging 45.

[107] Zie: F.J. Zuiderveen Borgesius, *Improving Privacy Protection in the area of Behavioural Targeting*, Alphen aan den Rijn: Kluwer Law International 2015, p. 240-241.

[108] Overweging 38 en 58a LIBE Compromis, proposal for a Data Protection Regulation. Deze bijdrage verwijst naar de Inofficial Consolidated Version after LIBE Committee Vote, provided by the Rapporteur, General Data Protection Regulation, 22 oktober 2013, www.janalbrecht.eu/fileadmin/material/Dokumente/DPR-Regulation-inofficial-consolidated-LIBE.pdf.

[109] Zie bijvoorbeeld G.J. Zwenne, 'Nog enkele opmerkingen over IP-adressen en persoonsgegevens, identificeerbaarheid en "single out"', *P&I* 2015, afl. 6, p. 216-221.

[110] Zie bijvoorbeeld R. Leenes, 'Do they know me? Deconstructing identifiability', *University of Ottawa Law and Technology Journal* (2008) 4(1-2) 135; Traung 2010; J. Koëter, 'Behavioral targeting en privacy: een juridische verkenning van internet gedragsmarketing', *IR* 2009, afl. 4, p. 104-111, m.n. p. 109. Zie ook P. De Hert & S. Gutwirth, 'Regulating profiling in a democratic constitutional state', in: M. Hildebrandt & S. Gutwirth (red.), *Profiling the European Citizen*, New York: Springer 2008, p. 288-289.

Ook maakt behavioural targeting discriminatie mogelijk. Bedrijven kunnen mensen indelen in 'targets' en 'waste', en hen zo behandelen.[114] Zo kan een adverteerder kortingen gebruiken om welvarende mensen te verleiden om klant te worden – en armere mensen uitsluiten van een campagne. Cookieprofielen kunnen gebruikt worden voor discriminatie, bijvoorbeeld als iemand (via een cookie) wordt geclassificeerd als 'disabled/handicapped',[115] of als 'lesbian, gay, bisexual, transgender'[116]. Er zijn dus veel risico's, ook als bedrijven geen namen

---

[111] Het HvJ EU merkt op dat surveillance (voor staatsveiligheid) een chilling effect kan veroorzaken: 'Bovendien kan het feit dat de gegevens worden bewaard en later worden gebruikt zonder dat de abonnee of de geregistreerde gebruiker hierover wordt ingelicht, bij de betrokken personen het gevoel opwekken dat hun privéleven constant in de gaten wordt gehouden (…)', HvJ EU 8 april 2014, gevoegde zaken C-293/12 en C-594/12, ECLI:EU:C:2014:238, r.o. 37 (*Digital Rights Ireland Ltd*). Zie F.J. Zuiderveen Borgesius & A. Arnbak, 'New Data Security Requirements and the Proceduralization of Mass Surveillance Law after the European Data Retention Case', Amsterdam Law School Research Paper No. 2015-41, http://ssrn.com/abstract=2678860, p. 19.

[112] Zie N.M. Richards, 'Intellectual privacy', (2008) 87 *Texas Law Review* 387. Zie ook: 'Zembla' (Data: het nieuwe goud II), www.npo.nl/zembla/09-12-2015/VARA_101375749.

[113] M. Marquis-Boire, G. Greenwald & M. Lee, 'XKEYSCORE: NSA's Google For The World's Private Communications' (The Intercept, 1 juli 2015), https://firstlook.org/theintercept/2015/07/01/nsas-google-worlds-private-communications/.

[114] De termen zijn van: J. Turow, *The Daily You: How the New Advertising Industry is Defining Your Identity and Your Worth*, New Haven: Yale University Press 2011.

[115] Rocket Fuel, Health Related Segments, http://rocketfuel.com/downloads/Rocket%20Fuel%20Health%20Segments.pdf.

[116] Flurry Audiences, Segment audiences by real interests, www.flurry.com/flurry-personas.html. Op 5 oktober 2013 noemde het bedrijf op deze website nog de categorie 'transgender'. Inmiddels noemt het bedrijf die

koppelen aan individuele profielen.

Een tweede reden om 'single-out'-gegevens als persoonsgegevens te zien is dat een naam slechts een van de identifiers is die kan worden gekoppeld aan gegevens van een persoon. Soms is een naam de meest praktische identifier. Als een bedrijf klantgegevens van een supermarkt-klantenkaart aan een cookieprofiel wil koppelen, zou het handig zijn als in beide datasets (cookies en supermarktgegevens) de individuele profielen aan een naam gekoppeld zijn. Maar in veel gevallen is een naam niet de meest praktische identifier. Als het doel is berichten versturen aan een telefoon, of de locatie van een telefoon bijhouden, is een telefoonnummer of een andere telefoon-ID een betere identifier dan een naam. Verder is een uniek nummer vaak een betere identifier dan een naam, omdat de namen niet uniek zijn.[117] Voor een advertentienetwerk dat iemands surfgedrag wil volgen, of advertenties wil richten op een persoon, is een cookie een betere identifier dan een naam.

Een derde reden waarom 'single-out'-gegevens als persoonsgegevens beschouwd moeten worden is dat het doel van behavioural targeting nu juist is om advertenties op *individuen* te richten. In de woorden van een marketeer: 'Zo wordt de juiste boodschap aan de juiste persoon op het juiste moment getoond.'[118] Zoals de International Working Group on Data Protection in Telecommunications opmerkt, willen adverteerders mensen beïnvloeden, en niet apparaten: 'While ads may well be addressed to a machine at the technical level, it is not the machine which in the end buys the proverbial beautiful pair of red shoes – it is an individual.'[119]

Ten vierde past het bij de ratio achter het gegevensbeschermingsrecht om naamloze behavioural-targeting-profielen als persoonsgegevens te zien. Een van de doelstellingen van het gegevensbeschermingsrecht is het beschermen van privacy en andere grondrechten.[120] Het Hof van Justitie van de Europese Unie zegt dat de richtlijn naar een 'hoog niveau' van

---

categorie niet meer. .

[117] De werkgroep merkt op dat veelvoorkomende namen niet altijd als persoonsgegevens gezien moeten worden (Artikel 29-Werkgroep 2007, WP 136, p. 13).

[118] Fingerspitz, Display advertising, www.fingerspitz.nl/diensten/display-advertising.

[119] International Working Group on Data Protection in Telecommunications, 'Web Tracking and Privacy' (juli 2013), www.datenschutz-berlin.de/attachments/949/675.46.13.pdf, p. 3.

[120] Artikel 1 en 6 lid 1 onderdeel a Richtlijn Bescherming Persoonsgegevens.

bescherming streeft,[121] en dat de richtlijn moet worden uitgelegd in het licht van grondrechten.[122] Bovendien moeten beperkingen met betrekking tot de bescherming van persoonsgegevens slechts gelden voor zover dat strikt noodzakelijk is.[123] Volgens het Europees Hof voor de Rechten van de Mens is het recht op privacy een brede term die dynamisch en pragmatisch moet worden toegepast.[124] Het is logisch om het gegevensbeschermingsrecht, net als het Europees Verdrag tot bescherming van de rechten van de mens, dynamisch uit te leggen. In het licht van nieuwe ontwikkelingen zoals behavioural targeting, profiling, en het Internet of Things, zou het niet zinvol zijn om de reikwijdte van gegevensbescherming te beperken tot gegevens die aan namen gekoppeld kunnen worden.

### 8. Argumenten tegen de 'single-out'-interpretatie

In de literatuur worden ook nadelen genoemd van een ruime interpretatie van het begrip 'persoonsgegeven'.[125] Een aantal van de belangrijkste argumenten tegen een ruime interpretatie wordt hier samengevat.

Ten eerste wordt aangevoerd dat bedrijven minder prikkels hebben om te investeren in pseudonimiseringstechnologie als de wet ook van toepassing is op pseudonieme gegevens.[126] Dat zou waar kunnen zijn. Maar ook als pseudonieme gegevens binnen de regels vallen, is er nog steeds een prikkel om gegevens te pseudonimiseren.

Op grond van het gegevensbeschermingsrecht en jurisprudentie van het Hof van Justitie van de Europese Unie, moeten verantwoordelijken persoonsgegevens beveiligen.[127]

---

[121] HvJ EU 16 december 2008, C-524/06, ECLI:EU:C:2008:724, r.o. 50 (*Huber*); HvJ EU 13 mei 2014, C-131/12, ECLI:EU:C:2014:317, r.o. 66 (*Google Spain*).

[122] Zie HvJ EU 13 mei 2014, C-131/12, ECLI:EU:C:2014:317, r.o. 68 (*Google Spain*).

[123] Zie bijvoorbeeld HvJ EU 8 april 2014, gevoegde zaken C-293/12 en C-594/12, ECLI:EU:C:2014:238, r.o. 52 (*Digital Rights Ireland Ltd*); HvJ EU 7 november 2013, C-473/12, ECLI:EU:C:2013:715, r.o. 39 (*Institut professionnel des agents immobiliers*) (met verdere verwijzingen).

[124] EHRM 11 juli 2002, 28957/95 (*Christine Goodwin/United Kingdom*), par. 74.

[125] Zwenne geeft een uitgebreid overzicht van argumenten tegen een brede interpretatie van het begrip 'persoonsgegeven': Zwenne 2013.

[126] G.J. Zwenne, 'Over Persoonsgegevens en IP-adressen, en de Toekomst van Privacywetgeving', in: L. Mommers e.a. (red.), *Het Binnenste Buiten. Liber Amicorum ter Gelegenheid van het Emeritaat van Prof. dr. Aernout H.J. Schmidt, Hoogleraar Recht en Informatica te Leiden*, Leiden: eLaw@Leiden 2010, p. 336.

[127] Artikel 17 Richtlijn Bescherming Persoonsgegevens; HvJ EU 8 april 2014, gevoegde zaken C-293/12 en C-

Pseudonimiseren kan de beveiliging van persoonsgegevens verbeteren. Stel bijvoorbeeld dat een datalek optreedt: een behavioural-targeting-bedrijf puliceert per ongeluk miljoenen naamloze cookie-profielen op het web. Men kan nu zien dat www.gevoelige-website.nl of www.voor-dit-onderwerp-schaam-ik-me.nl is bezocht door de persoon achter cookie *xyz*. Maar iemand die deze informatie vindt, ziet niet direct de naam van de persoon die deze websites heeft bezocht. De privacyrisico's voor de persoon achter cookie *xyz* zijn dus verminderd, omdat het datalek naamloze profielen betreft. Maar een naam vervangen door een andere identifier is niet genoeg om persoonsgegevens te anonimiseren, of om de gegevens buiten het toepassingsgebied van het gegevensbeschermingsrecht te houden.[128]

Ten tweede wordt wel gesuggereerd dat het slecht zou zijn voor de economie en innovatie als pseudonieme gegevens onder het gegevensbeschermingsrecht zouden vallen.[129] Zelfs als dat zou kloppen, dan is dat argument niet genoeg om pseudonieme gegevens buiten het toepassingsgebied van het gegevensbeschermingsrecht te houden. Als gegevens onder het gegevensbeschermingsrecht vallen, dan betekent dat nog niet dat de verwerking is verboden, of dat de gegevens alleen met de toestemming van de betrokkene verwerkt mogen worden. Het betekent wel dat bedrijven moeten voldoen aan het gegevensbeschermingsrecht, en dus op een behoorlijke manier moeten omgaan met de gegevens.

Sommige bedrijven zouden minder winst maken als ze moesten voldoen aan het gegevensbeschermingsrecht. Het gegevensbeschermingsrecht eist bijvoorbeeld dat persoonsgegevens goed beveiligd worden: dat kost geld. Maar zelfs als we fundamentele rechten zouden negeren en alleen op economische effecten zouden letten, zou een meer relevante vraag zijn of de samenleving er als geheel op voor- of achteruitgaat als bedrijven moeten voldoen aan het gegevensbeschermingsrecht. Acquisti, de toonaangevende onderzoeker op het gebied van de economische aspecten van privacy, toont aan dat het onduidelijk is of meer of minder wettelijke bescherming van persoonsgegevens beter is vanuit een economisch perspectief: '(...) economic theory shows that, depending on conditions and assumptions, the protection of personal privacy can increase aggregate welfare as much as the

---

594/12, ECLI:EU:C:2014:238, r.o. 66 (*Digital Rights Ireland Ltd*). Zie over beveiliging ook: A. Arnbak, *Securing Private Communications* (diss. Amsterdam UvA), 2015, http://hdl.handle.net/11245/1.492674.
[128] Artikel 29-Werkgroep 2014, WP 216.
[129] Zie Stringer 2013.

interruption of data flows can decrease it.'[130]

Bovendien: innovatie en economische groei zijn belangrijk. Maar die doelen wegen niet zwaarder dan grondrechten.[131] Als het goed zou zijn voor innovatie of de economie als kinderen onder de acht jaar in fabrieken zouden werken, zouden we dat toch niet toestaan. Afgezien daarvan: als de wet bedrijven ertoe beweegt om privacyvriendelijke behavioural-targeting-technieken te ontwikkelen, dan is dat ook innovatie.[132]

Ten derde wordt wel gezegd dat een ruime interpretatie van persoonsgegevens ertoe zou leiden dat het gegevensbeschermingsrecht ook van toepassing is als privacy niet wordt bedreigd. Sommigen suggereren dat het recht op bescherming van persoonsgegevens niet losgekoppeld zou moeten worden van het recht op privacy.[133] Dit argument is moeilijk te rijmen met het geldende recht. Het Handvest van de grondrechten van de Europese Unie onderscheidt het recht op bescherming van persoonsgegevens en het recht op privacy.[134] Bovendien zeggen veel auteurs dat het juist een voordeel is dat het gegevensbeschermingsrecht geldt voor alle persoonsgegevens, in plaats van alleen voor privacygevoelige gegevens.[135]

Ten vierde zijn sommigen bang dat bijna alles een persoonsgegeven zou kunnen worden als het begrip 'persoonsgegeven' te ruim wordt geïnterpreteerd. Handhaving zou te moeilijk worden. Privacytoezichthouders zouden het recht alleen kunnen handhaven tegen sommige overtreders. Dit kan leiden tot arbitraire beslissingen over handhaving. Dat zou slecht zou zijn voor de rechtszekerheid.[136] Een verwant punt is dat de reikwijdte van het begrip 'persoonsgegevens' te onzeker zou worden. Ook dat zou slecht zijn voor de rechtszekerheid. Er zit wat in dat een brede interpretatie van het begrip 'persoonsgegeven' handhaving van het gegevensbeschermingsrecht moeilijk maakt. Maar de reikwijdte van het begrip 'persoonsgegeven' inperken zou niet de juiste reactie zijn. Ter vergelijking: het is waarschijnlijk goed dat we milieurecht hebben, ook al is het onmogelijk om alle overtreders te betrappen. Bovendien kan in de juridische praktijk een definitie altijd tot discussie leiden. Alleen ervoor zorgen dat het gegevensbeschermingsrecht van toepassing is op behavioural targeting is niet voldoende om alle privacyproblemen van behavioural targeting op te lossen. Het gegevensbeschermingsrecht heeft zwakke punten, en naleving en handhaving laten te wensen over. Maar het gegevensbeschermingsrecht biedt wel een kader om te beoordelen of persoonsgegevens zorgvuldig en behoorlijk worden verwerkt. Bovendien kan het gegevensbeschermingsrecht helpen om gegevensverwerkingen transparant te maken, omdat het bedrijven verplicht om informatie openbaar te maken over hun gegevensverwerkingen.[137] Als problemen aan het licht komen, zou dit kunnen leiden tot de conclusie dat meer regelgeving nodig is.

---

[130] A. Acquisti, 'The economics of personal data and the economics of privacy' (background paper conference: The Economics of Personal Data and Privacy: 30 Years after the OECD Privacy Guidelines) (2010), www.oecd.org/internet/ieconomy/46968784.pdf, p. 19. Zie ook: A. Acquisti, 'The Economics and Behavioral Economics of Privacy', in: J. Lane e.a. (red.), *Privacy, Big Data, and the Public Good: Frameworks for Engagement*, Cambridge: Cambridge University Press 2014, p. 90.

[131] Zie ook: H. Hijmans 2016, p. 315 (hoofdstuk 6, par. 9).

[132] Amerikaanse onderzoekers hebben bijvoorbeeld een behavioural-targeting-systeem ontwikkeld waarbij alle persoonsgegevens worden opgeslagen in de computer van de betrokkene; zo wordt de privacyinbreuk geminimaliseerd. V. Toubiana e.a., 'Adnostic: Privacy Preserving Targeted Advertising', 2010, http://crypto.stanford.edu/adnostic.

[133] C. Cuijpers & P. Marcelis, 'Oprekking van het Concept Persoonsgegevens Beperking van Privacybescherming?', *Computerrecht* 2012, afl. 6, 339-351.

[134] Artikel 7 en 8 Handvest van de grondrechten van de Europese Unie.

[135] Zie bijvoorbeeld P. De Hert & S. Gutwirth, 'Privacy, Data Protection and Law Enforcement. Opacity of the Individual and Transparency of Power', in: E. Claes, A. Duff & S. Gutwirth (red.), *Privacy and the Criminal Law*, Antwerpen: Intersentia 2006, p. 94. Zie voor een andere benadering van de verhouding tussen het gegevensbeschermingsrecht en het recht op privacy: Hijmans 2016, p. 66-71 (hoofdstuk 2, par. 13).

## 9. Conclusie

Er is veel discussie over de vraag of het gegevensbeschermingsrecht van toepassing is als bedrijven gegevens over mensen verwerken, maar daar geen namen aan koppelen. Voor behavioural targeting worden bijvoorbeeld naamloze profielen verzameld van honderden miljoenen mensen.

Veel privacyrisico's blijven bestaan, ongeacht of bedrijven een naam koppelen aan de informatie over een individu. Zo kan grootschalige dataverzameling een *chilling effect* veroorzaken, ook als bedrijven werken met naamloze (pseudonieme) gegevens over mensen. En een cookieprofiel dat zegt dat iemand gehandicapt is of uit een arme buurt komt, kan gebruikt worden voor discriminatie. Bovendien is een naam slechts een van de identifiers die aan gegevens over een persoon gekoppeld kunnen worden, en voor behavioural targeting niet eens de meest praktische identifier. Kortom, gegevens die gebruikt worden om iemand te onderscheiden binnen een groep, *to single out somebody*, moeten als persoonsgegevens beschouwd worden – ook als bedrijven geen naam aan die gegevens koppelen. Bovendien is het vaak vrij eenvoudig voor bedrijven om een naam te verbinden aan pseudonieme gegevens. Twee conclusies kunnen worden getrokken. Ten eerste laat een analyse van het geldende recht, in ieder geval volgens de interpretatie van de Europese privacytoezichthouders, zien dat de regels voor persoonsgegevens doorgaans van toepassing zijn op behavioural-targeting-gegevens. Ten tweede zou het gegevensbeschermingsrecht ook vanuit een normatief perspectief van toepassing moeten zijn.

---

[136] Zwenne 2013, p. 13-16.

[137] Artikel 10 en 11 Richtlijn Bescherming Persoonsgegevens.

\* \* \*